\documentclass{llncs}
\usepackage{epsfig}
\usepackage{graphicx}
\begin{document}
\mainmatter              
\def\dom{\mathcal{D}}
\def\diag{\mathrm{diag}}
\def\trace{\mathrm{tr}}
\def\inv{^{-1}}
\def\adj{^\dagger }
\def\invsq{^{-\frac 1 2 }}
\def\degree{^\circ} 
\newcommand\mypar[1]{\medskip\par\noindent\textbf{#1}\ \ }
\title{Source separation on astrophysical data sets from the WMAP satellite}
\titlerunning{ICA on WMAP}  
%
\author{G. Patanchon\inst{1}, J. Delabrouille\inst{2} \and J.-F. Cardoso\inst{3}}
\authorrunning{}   

\tocauthor{
G Patanchon (University of British Columbia)
J Delabrouille (CNRS/PCC, P7/APC)
JF Cardoso (CNRS/LTCI, P7/APC),
}
\institute{University of British Columbia, Canada
\and
CNRS/PCC, UMR 7553, Paris, France and APC, Paris France
\and
CNRS/LTCI, UMR 5141,  Paris, France \texttt{http://tsi.enst.fr/\homedir cardoso}
}
\maketitle              

\begin{abstract}
  This paper presents and discusses the application of blind source separation
  to astrophysical data obtained with the WMAP satellite. Blind separation permits
  to identify and isolate a component compatible with CMB emission, and to measure
  both its spatial power spectrum and its emission law. Both are found to be compatible 
  with the present concordance cosmological model. This application demonstrates 
  the usefulness of blind ICA for cosmological applications.
\end{abstract}

\def\bR{\mathbf{R}}
\def\bA{\mathbf{A}}
\def\bS{\mathbf{S}}
\def\bD{\mathbf{D}}
\def\bN{\mathbf{N}}
\def\bX{\mathbf{X}}
\def\bX{\mathbf{X}}
\def\bW{\mathbf{W}}
\def\bC{\mathbf{C}}

\def\adj{^\dagger}
\def\inv{^{-1}}
\def\invadj{^{-\dagger}}
\def\diag{\mathrm{diag}}
\def\vdiag{\mathrm{vdiag}}
\def\trace{\mathrm{tr}}

\newcommand\kull[2]{D(#1, #2)}

\section{Introduction}

The detection of Cosmic Microwave Background (CMB) fluctuations, and the measurement of their spatial power spectrum, has been over the past three decades subject of intense activity in the cosmology community. 

The CMB, discovered in 1965 by Penzias and Wilson, is a relic radiation emitted some 13 billion years ago, when the Universe was about 370.000 years old. This radiation emits, as a function of wavelength, as an almost perfect blackbody at a temperature of 2.726 Kelvin. Small fluctuations of this temperature tracing the seeds of the primordial homogeneities which gave rise to present large scale structures as galaxies and clusters of galaxies, are expected to exist, at the level of a few tens of micro Kelvin.

The importance of measuring these fluctuations to constrain cosmological scenarios describing the history and properties of our Universe is now well 
established.  The measurement of their statistical properties (and in particular of their spatial power spectrum) permits to drastically constrain the cosmological parameters describing the matter 
content, the geometry, and the evolution of our Universe
\cite{jungman96}.  The accuracy required for precision 
tests of the cosmological scenarios, 
however, is such that it is necessary to disentangle in the data the 
contribution of several distinct astrophysical sources, all of which 
emit radiation in the frequency range used for CMB observations \cite{fb-rg99}. In addition, all measurements being noisy, it is necessary to identify, characterize, and remove as well as possible contributions due to instrumental noise.

CMB emission peaks in the millimeter-wave domain of the electromagnetic spectrum. In this wavelength range however, several astrophysical sources of radiation contribute to the total emission. Some of them are quite well known (or modelled), and others not so well. It is of utmost importance then, for the interpretation of the observations, to separate the various emissions. This can be achieved (to some extent) by the joint processing of various observations performed at various wavelengths.

To first order, the observation at a given wavelength $\lambda$ can be modeled as a linear superposition of a number of sources.
A sequence $\{\bX(x,y)\}$ of observations  --- $J$ maps obtained at several wavelengths --- is modeled as an
instantaneous mixture of $K$ independent sources contaminated
by independent noise $\bN(x,y)$:
\begin{displaymath}
  \bX(x,y) = \bA  \bS(x,y) + \bN(x,y)
\end{displaymath}
where $\bA$ is an unknown $J\times K$ matrix.  

As a strong prediction of the standard big bang model of cosmology,  
CMB anisotropies are expected to have a spectral emission law given, to first order, by the derivative (with respect to temperature) of the blackbody law, taken at the temperature of the CMB:
\begin{equation}
\Delta I = \Delta T \times \left[ \frac{\partial B_\nu(T)}{\partial T} \right]_{T=2.726\,{\rm K}} 
\end{equation}
where $B_\nu(T)$ is the Planck law for the emission of a blackbody at temperature $T$.
Blind component separation, which permits to estimate the emission law of sky components, provides a unique tool for checking this prediction.
In the following, we apply a blind component separation method on WMAP data.

\section{WMAP data}

\subsection{WMAP}

The WMAP space probe, launched by NASA in 2001, is a large telescope for imaging the total emission of the sky at 5 different wavelengths (or frequency channels), with a resolution ranging from about 0.2 degrees to 0.9 degrees (limited by diffraction), and with full sky coverage ($4\pi$ of the sphere) \cite{map-basic}.

Data taken with the WMAP instrument has been made available to the scientific community after one year of proprietary period, and is freely available on a dedicated web site.\footnote{http://lambda.gsfc.nasa.gov/} In its most usable format, the data consists of a set of 10 maps obtained by different detector pairs: four maps at 3.2 mm (94 GHz), two at 4.9 (61 GHz), and one at 7.3, at 9.1, and at 13.0 mm (41, 33 and 23 GHz). The data is provided in the Healpix pixellisation format of the sphere.\footnote{http://www.eso.org/science/healpix/}
Data from the various detectors observing the sky at the same frequency can be combined to yield a single map of the sky at that frequency. In this analysis, we use the data from individual detector pairs.

\subsection{Astrophysical emissions}

In the millimeter waveband, the total sky emission can be considered as the superposition of two main classes of sources: diffuse emissions and compact sources.

Diffuse emissions are due to extended processes on the sky. For all of these processes,  the observed photons are emitted by large scale distributions of emitters -- gases of particles. Among these, one expects emission from the hot plasma in the early Universe (the CMB), greybody emission from cold dust particles in the galaxy (dust emission), emission from relativistic electrons spiraling in the galactic magnetic field (synchrotron), free--free (Bremsstrahlung) emission from ionized gaz.
Compact source emission is due to distant objects (galaxies, quasars, clusters of galaxies), the angular size of which is smaller than the resolution of the instrument.

For each physical source of emission, the total observable intensity $I(\lambda, \theta, \phi)$, depends on the physical properties of the source (clumpiness, emission process, temperature, etc...). Here, $\lambda$ is the wavelength and $(\theta,\phi)$ is the direction on the celestial sphere.
For most diffuse emissions, we can approximate
$I(\lambda, \theta, \phi) \simeq \bA(\lambda) \bS(\theta,\phi)$, 
the approximation being expected to be excellent (less than 0.1 \%) for some components (e.g. CMB) and not so good (at the 10\% level) for others as dust emission.

\subsection{Spherical harmonics}

It is convenient and usual to decompose the sky brightness (traditionnally expressed as $\Delta T(\theta, \varphi)$ in units of temperature fluctuation in micro Kelvin), as an expansion:
\begin{equation}
\Delta T(\theta, \varphi) = \sum_{\ell,m} a_{\ell,m} Y_{\ell,m}(\theta, \varphi)
\end{equation}
where the functions $Y_{\ell,m}(\theta, \varphi)$ are the usual spherical harmonics. For a stationary isotropic random Gaussian field on the sphere, the coefficients $a_{\ell,m}$ on the sphere are such that
%
$\mathrm E (a_{\ell,m}a_{\ell',m'}) = C_\ell \delta_{\ell \ell'} \delta_{m m'} $.
On the spherical harmonics basis, a noisy linear mixture of components writes
\begin{equation}
  \bX(\ell,m) = \bA  \bS(\ell,m) + \bN(\ell,m)
  \label{eq:model-lm}
\end{equation}
The WMAP observations used in this application are decomposed on spherical harmonics, and the component separation method is applied on the set $\{ \bX(\ell,m)\}$, a $10 \times 1$ vector of coefficients of the spherical harmonics expansion of the WMAP observations.

\section{Method}

\subsection{Spectral Matching}

Following the method described in \cite{SM-mnras}, we compute average values of the covariance of the observations:
\begin{equation}
\widehat \bR_\bX(q) = \frac 1 {n_q} \sum_{(\ell,m) \in \dom_q}  \bX(\ell,m) \bX(\ell,m)\adj
  \ \ \ \ (q=1,\ldots, Q)
\end{equation}
where $q$ labels a domain $ \dom_q$ of $(\ell,m)$ space and $n_q = \mathrm{card}(\dom_q)$. For our application, it is customary to estimate band averages of the power spectra in domains $\ell_1 < \ell \leq \ell_2$, with $m$ taking all available values for each $\ell$, i.e. $-\ell \leq m \leq \ell$.

If the observations $\bX$ match the multi-component model of equation  \ref{eq:model-lm}, then 
$\widehat \bR_\bX(q)$ is an estimator of $\bR_\bX(q)$, the latter being itself given by 
\begin{equation}
  \label{eq:strucSCM}
  \bR_\bX(q) = \bA \bR_\bS(q) \bA\adj + \bR_\bN(q)
  \ \ \ \ (q=1,\ldots, Q)
\end{equation}
with $\bR_\bS(q)$ and $\bR_\bN(q)$ defined similarly to $\bR_\bX(q)$. The parameters of interest (mixing matrix and  band-averaged spatial power spectra of components) are precisely $\bA$ and $\bR_\bS(q)$, the latter being diagonal for uncorrelated components. Noise spatial power spectra, being of no particular use for astrophysics, are nuisance parameters which must be estimated, so that the full set of parameters of the data is $\theta = \{\bA,\bR_\bS,\bR_\bN\}$.

The estimation of these parameters of the model is done by maximizing an approximation of the likelihood (in the Whittle approximation), which (up to irrelevant factors) boils down to minimizing
\begin{equation}\label{eq:obj}
  \phi(\theta)
  =
  \sum_{q=1}^Q
  n_q
  \
  D\left(\widehat \bR_\bX(q),  \bR_\bX(q; \theta) \right) 
\end{equation}
$D(\cdot,\cdot)$, the Kullback-Leibler divergence, is a measure of divergence between two positive
$n\times n$ matrices, defined by
\begin{equation}
  \label{eq:kullR}
  D(\bR_1,\bR_2) 
  = \frac{1}{2} \left [
  \trace \left( \bR_1\bR_2\inv \right) - \log\det (\bR_1\bR_2\inv) - n  \right ]
\end{equation}

Optimization is achieved with a dedicated algorithms based on the EM algorithm, followed by few steps of a descent method (BFGS algorithm) to speed-up the convergence \cite{cardosopham}.

\subsection{Estimation of Component maps}

Component maps are estimated by applying to the observations $\bX$ a Wiener filter 
$\bW = [\bA^t \bR_\bN^{-1} \bA + \bR_\bS^{-1}]^{-1}\bA^t \bR_\bN^{-1}$, where estimated values $\widehat \bA$, $\widehat \bR_\bN$ and $\widehat \bR_\bS$ of the parameters are used. The estimated components are obtained then by:
\begin{equation}
\widehat \bS = [\widehat \bA^t \widehat \bR_\bN^{-1} \widehat \bA + \widehat \bR_\bS^{-1}]^{-1}\widehat \bA^t \widehat \bR_\bN^{-1} \bX
\end{equation}

In the limiting case where noise is small as compared to component signals, $\bR_\bS^{-1}$ is negligible. If in addition estimates of the mixing matrix is good ($\widehat \bA \simeq \bA$) the application of the Wiener filter to the observations yields
\begin{equation}
\widehat \bS \simeq \bS + [\bA^t \bR_\bN^{-1} \bA]^{-1}\bA^t \bR_\bN^{-1} \bN
\end{equation}
which is 
an unbiased estimate of $\bS$. In poor signal to noise regimes, the signal is attenuated to suppress noise contamination in the reconstructed maps.

\subsection{Dealing with real data}

The observations of astrophysical emissions are more complicated than the simple model of equation \ref{eq:model-lm} in several respects.
\begin{itemize}
\item Known galactic emissions (dust, synchrotron, free-free) are positive and concentrated towards the galactic plane;
\item In addition, galactic component emissions may be correlated for physical reasons (thus violating the independence assumption);
\item The emission law (mixing matrix elements) of galactic components is known to be slightly position dependent and/or scale dependent;
\item Point sources are not modeled satisfactorily by a linear mixture, and their distribution is poorly represented by the Whittle approximation;
\item Instrumental noise is not stationary over the sky, which reduces somewhat the efficiency of parameter estimation;
\item Detectors observe the sky with a detector dependent angular resolution. This effect can be modeled, to first order, as a detector-dependent multiplicative term $B_\ell$ in harmonic space.
\end{itemize}

For the present analysis, we deal with these complications in the simplest possible way.
We decide to focus the present analysis on checking whether WMAP data contains a component which has both a CMB power spectrum and the expected emission law for temperature fluctuations of the cosmological blackbody. Hence, we cut out from the analysis those regions of the sky which are most contaminated by galactic emission. Also, we cut out from the analysis small regions around known bright point sources (see next subsection). This reduces drastically the impact of the first four items in the list. We ignore noise non stationarity. Irregularities are smooth enough on the sky to induce only small (and small range) correlations in harmonic space. The impact of the multiplicative term $B_\ell$ in harmonic space is corrected for beforehand.

We restrict the analysis to multipole $\ell$ values between 10 and 400, which encompasses all of the first acoustic peak observed by WMAP using the highest frequency channels only. The upper limit permits to minimize the impact of a background of residual unresolved point sources and of uncertainties in the modeling of the beam of the instrument. 

\subsection{Mask}

We apply a mask to the observations which sets to 0 regions of the sky contaminated by strong emission which probably are not well modeled as a linear mixture of uncorrelated sources. These regions are essentially a set of small cuts centered on known bright point sources, and a large cut through the map to exclude the galactic plane region, highly contaminated by a complex set of correlated foreground emissions. This cut permits to isolate a region (the major fraction of the sky) where the emission is thought to be dominated by CMB fluctuations. 

It is possible to perform the SMICA analysis using the decomposition in spherical harmonics of cut---sky data.
The analysis, then, is performed on a fraction of the sky (at galactic latitude higher than 20 degrees, with point sources removed, and with small additional high--level galactic emission regions cut). 

On such incomplete sky, some technical issues have to be taken into account. First, the spherical harmonics do not form an orthonormal basis, but his has little impact on the estimation in large bands of $\ell$. 
Second, because of the mask, the stationarity hypothesis which guarantees the decorrelation of spherical harmonics expansion coefficients of the components is not quite satisfied, which results in possible loss of efficiency. 
As a final comment on masking, we note that the spectra estimated on the cut sky have to be corrected to take into account the impact of this ``window function". We apply a correction using the Master formalism of Hivon et al. \cite{master} to correct from partial sky coverage effects.

\section{Results}

The application of SMICA to WMAP data gives the following results:

\begin{itemize}
\item Two components are clearly identified, one of which is CMB, and the second emission from the high latitude galaxy.
\item A third component is identified marginally, which may correspond to large scale variations of galactic emission law,
\item The identified CMB component has an emission law compatible with the expected derivative of a blackbody, to excellent accuracy. Its power spectrum displays a peak around $\ell = 200$ (first acoustic peak) compatible with the measurement announced by the WMAP team.
\item The galactic component has an emission law compatible with synchrotron emission in all WMAP channels (proportional to $\nu^{-2.7}$) except those at 90 GHz, for which a significant excess is seen. This excess is interpreted as due to galactic dust emission correlated with the synchrotron (either for physical reasons or -- more likely -- by happenstance because of concentration of galactic emissions at low galactic latitude).
\end{itemize}

Error bars for all quantities are computed from the Hessian at the point of convergence, with a correction factor $1/\sqrt(f_{sky}$ for part--sky coverage.

\begin{figure}[th]
\begin{center} 
\includegraphics[width=0.49\textwidth]{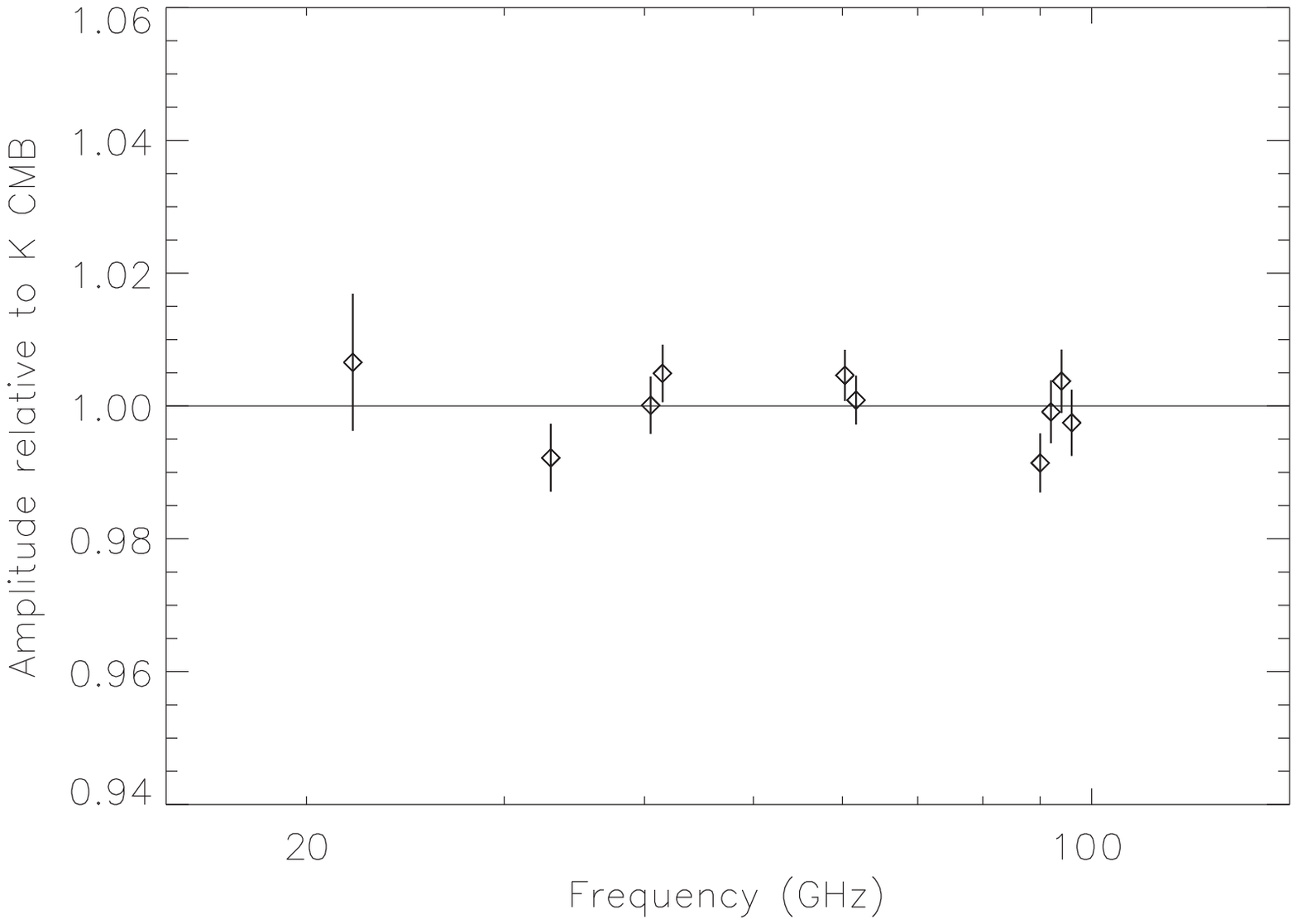}
\includegraphics[width=0.49\textwidth]{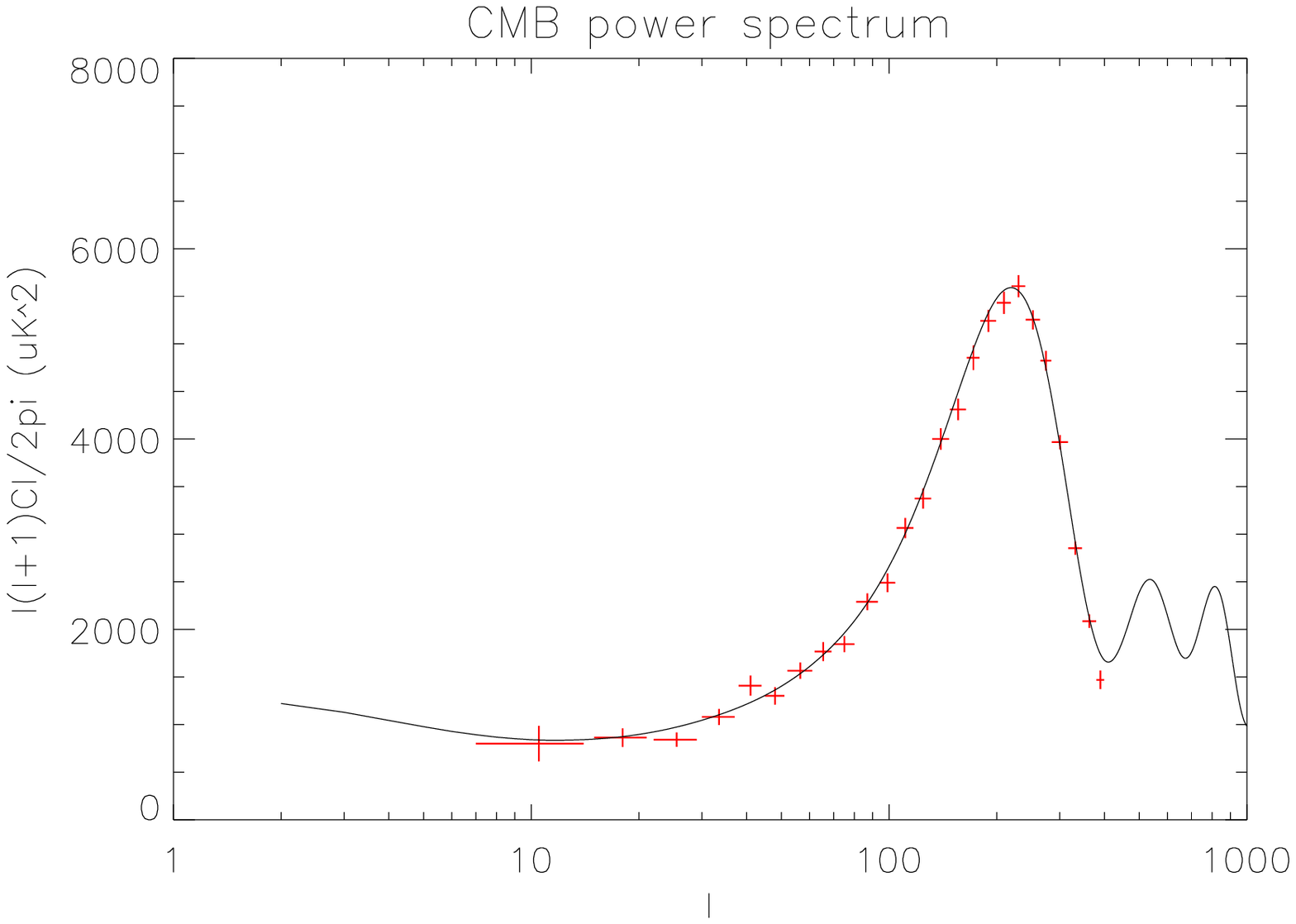}
\caption{\label{fig:CMB}Left : Measured CMB fluctuation mixing matrix, compared to theoretical 
expectations. Data points at the same frequency have been slightly offset for readability. This figure shows that the first component identified blindly in the WMAP data, and seen consistently in all WMAP channels, has a CMB fluctuation emission law to within less than one per cent (error bars here exclude WMAP calibration errors). Right: Measured CMB spatial power spectrum, compared to the theoretical model matching best WMAP team measurement (solid black line). This figure shows that the first component identified blindly in the WMAP data, and seen consistently in all WMAP channels, has a spatial power spectrum consistent with what has been measured by WMAP using only the two highest frequency channels.}
\end{center}
\end{figure}

\begin{figure}[b]
\begin{center} 
\includegraphics[width=7cm,angle=90]{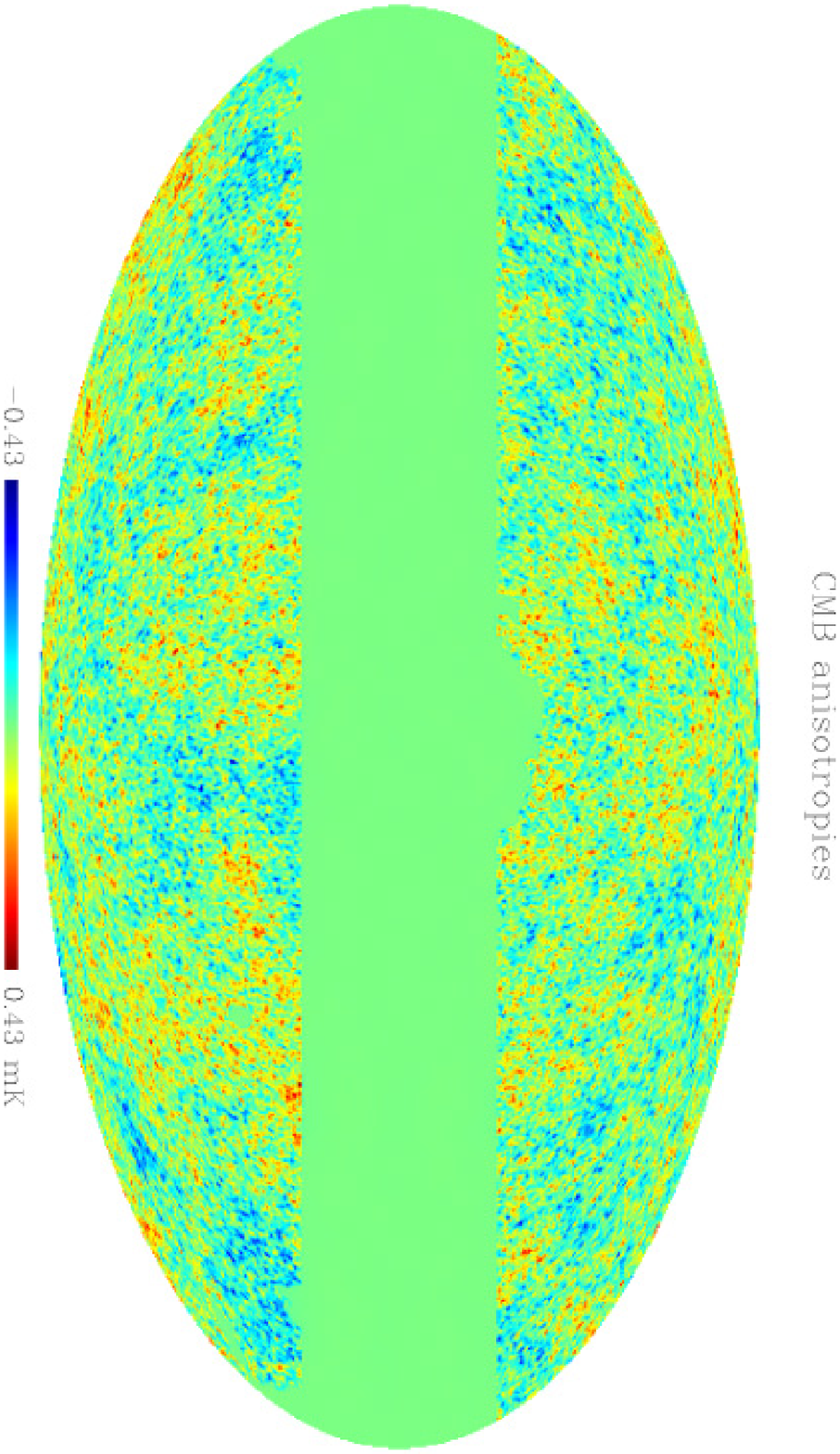}
\vspace{2cm}
\includegraphics[width=7cm,angle=90]{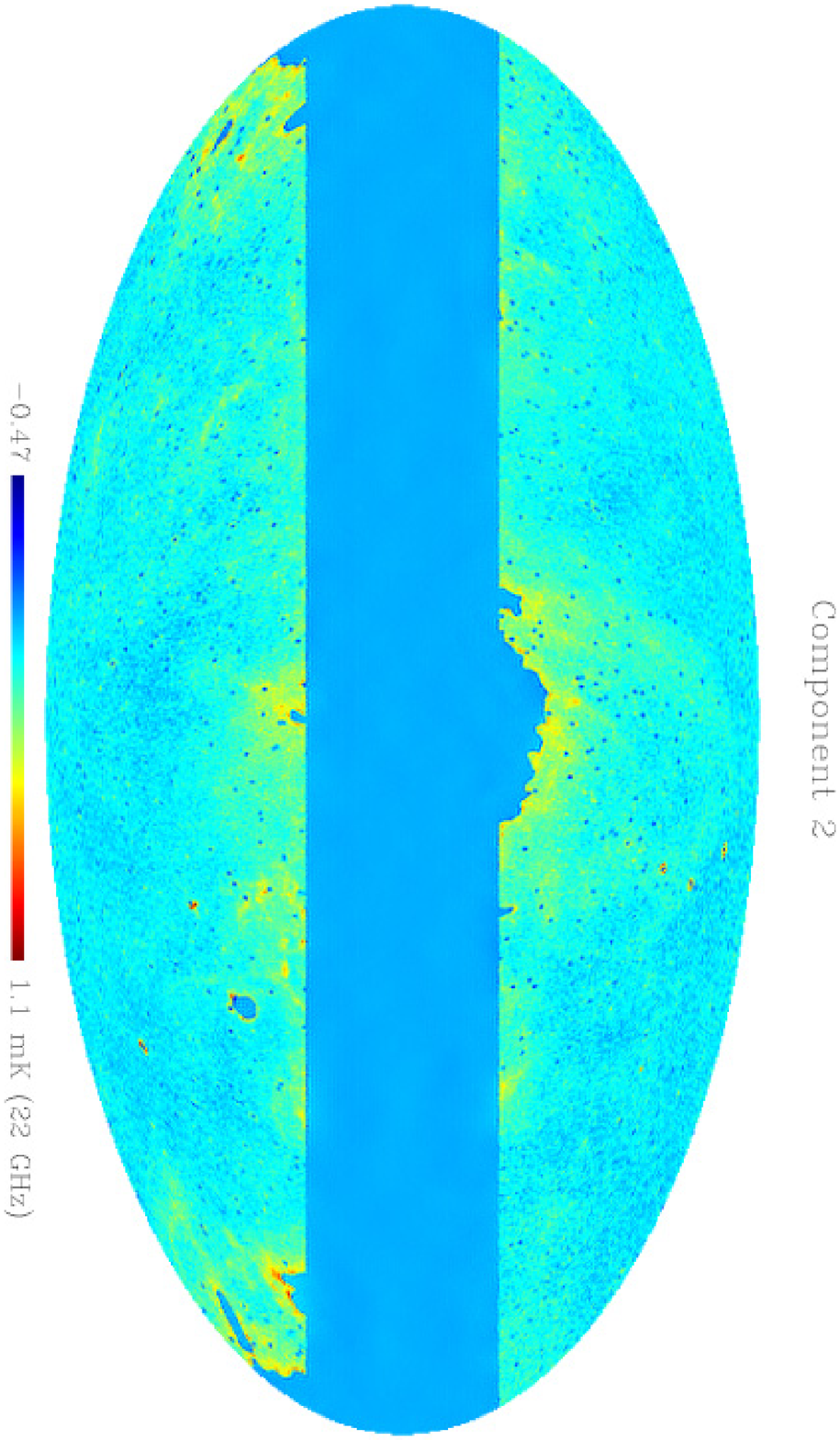}
\caption{\label{fig:CMB-map}Top: The map of CMB fluctuations obtained after Wiener filtering of the data using parameters estimated with SMICA. Bottom: The map of the second component (galactic emission) obtained in the same way.}
\end{center}
\end{figure}

\section{Conclusion} 

The SMICA blind component separation has been applied successfully to WMAP data. The analysis is performed on $\ell$ varying from about 10 to 400, which encompasses all of the first acoustic peak of the CMB. We find that all 10 WMAP maps comprise a common astrophysical component identified here as CMB. The emission law of this first component is compatible with CMB fluctuations to within about one per cent, and its spatial power spectrum in the range $10 \leq \ell \leq 400$ compatible with the measurement obtained by the WMAP team. The second component has an emission law proportional to $\nu^{-2.7}$ except in the W (90 GHz) channel for which a significant excess is observed. This excess is thought to be due to dust emission contribution to the component.
These results demonstrate the usefulness of blind component separation for astrophysical data analysis.

%
%

\end{document}